\documentclass[preprint,aps]{revtex4}

\usepackage{graphicx}
\usepackage{dcolumn}
\usepackage{bm}

\begin{document}



\title{Two-gap and paramagnetic pair-breaking effects on upper critical
field of SmFeAsO$_{0.85}$ and SmFeAsO$_{0.8}$F$_{0.2}$ single
crystals}

\author{Hyun-Sook Lee$^1$, Marek Bartkowiak$^2$, Jae-Hyun
Park$^1$, Jae-Yeap Lee$^1$, Ju-Young Kim$^3$, Nak-Heon Sung$^3$,
B. K. Cho$^3$, Chang-Uk Jung$^4$, Jun Sung Kim$^1$, and Hu-Jong
Lee$^1$}

\altaffiliation{Corresponding author: hjlee@postech.ac.kr}

\affiliation{$^1$ Department of Physics, Pohang University of
Science and Technology, Pohang 790-784, Republic of Korea \\
$^2$Hochfeld-Magnetlabor Dresden (HLD), Forschungszentrum
Dresden-Rossendorf, Dresden, Germany \\
$^3$Department of Materials Science and Engineering, GIST, Gwangju
500-712, Republic of Korea \\
$^4$Department of Physics, Hankuk
University of Foreign Studies, Yongin, Gyeonggi 449-791, Republic
of Korea}


\begin{abstract}
We investigated the temperature dependence of the upper critical
field [$H_{c2}(T)$] of fluorine-free SmFeAsO$_{0.85}$ and
fluorine-doped SmFeAsO$_{0.8}$F$_{0.2}$ single crystals by
measuring the resistive transition in low static magnetic fields
and in pulsed fields up to 60 T. Both crystals show that
$H_{c2}(T)$'s along the $c$ axis [$H_{c2}^c(T)$] and in an
$ab$-planar direction [$H_{c2}^{ab}(T)$] exhibit a linear and a
sublinear increase, respectively, with decreasing temperature
below the superconducting transition. $H_{c2}(T)$'s in both
directions deviate from the conventional one-gap
Werthamer-Helfand-Hohenberg theoretical prediction at low
temperatures. A two-gap nature and the paramagnetic pair-breaking
effect are shown to be responsible for the temperature-dependent
behavior of $H_{c2}^c$ and $H_{c2}^{ab}$, respectively.
\end{abstract}


\maketitle
\section{Introduction}
The upper critical field, $H_{c2}$, is one of the most important
superconducting parameters, providing a valuable insight into the
pairing mechanism and information on fundamental superconducting
properties such as coherence length scales. The temperature
dependence of the upper critical field $H_{c2}(T)$ and its
anisotropy are sensitive to the details of the underlying
electronic structures and reflect the dimensionality of
superconductivity. Furthermore, 
$H_{c2}$ is related to the critical current density, which is an
important material parameter for application purposes.

After the discovery of iron-pnictides-based
superconductors~\cite{Kamihara}, REFeAsO$_{1-x}$F$_x$ (RE =
rare-earth elements), with relatively high superconducting
transition temperature ($T_c$), many efforts have been made to
investigate their $H_{c2}(T)$. REFeAsO$_{1-x}$F$_x$ has a layered
structure with alternating stacks of insulating REO and conducting
FeAs layers. Despite the presence of the two-dimensional nature in
the materials, most of studies on $H_{c2}(T)$ of
REFeAsO$_{1-x}$F$_x$ have been limited to
polycrystals~\cite{Hunte,Jo} because of the difficulty with
growing REFeAsO$_{1-x}$F$_x$ single crystals. Therefore, recent
investigations of single crystals have been more focused on
AEFe$_2$As$_2$ (AE = alkaline-earth
elements)~\cite{Rotter,Ni,Sasmal} single crystals, which can be
grown with relative ease in the ambient conditions. High-field
measurements on electron- and hole-doped AEFe$_2$As$_2$ single
crystals and films~\cite{Altarawneh,Yamamoto,Baily,Yuan,Kano}
showed that $H_{c2}(T)$ in a $c$-axis field [$H_{c2}^c(T)$] and in
an $ab$-planar field [$H_{c2}^{ab}(T)$] increases almost linearly
and sublinearly with decreasing temperature below $T_c$,
respectively, regardless of the doping level and the degree of
disorder. The resultant $H_{c2}^{ab}(T)$ approaches the value of
$H_{c2}^{c}(T)$ at $T$ far below $T_c$, leading to the almost
isotropic superconductivity in the zero-temperature limit. Such a
quasi-isotropic property of $H_{c2}(T)$ in a layered structure is
quite intriguing and has been attributed to the multi-band
effect~\cite{Baily,Yuan,Kano}.

Recently single crystalline REFeAsO$_{1-x}$F$_x$ (RE = Sm, Nd, Pr)
~\cite{Zhigadlo,Prozorov,Jia1,Karpinski} were successfully grown
by using the flux-growth technique or the
high-pressure-high-temperature technique. There is, however, a
single report on the $H_{c2}(T)$ of REFeAsO$_{1-x}$F$_x$ single
crystals for NdFeAsO$_{0.7}$F$_{0.3}$ at temperatures far below
$T_c$, where $H_{c2}(T)$ was traced by high-field resistivity
measurements using pulsed magnetic field up to 60
T~\cite{Jaroszynski}. $H_{c2}^c(T)$ in the study exhibited a
pronounced upturn curvature at low
temperatures~\cite{Jaroszynski}. By contrast, $H_{c2}^{ab}(T)$
showed a downturn curvature in the low-temperature
range~\cite{Jaroszynski}. Apparently, the temperature dependence
of $H_{c2}^c$ for NdFeAsO$_{1-x}$F$_x$, despite being common
iron-arsenic compounds, appears to be quite different from that of
AEFe$_2$As$_2$. Thus, it is highly required to examine any common
or dissimilar $H_{c2}(T)$ behavior of different compounds of
REFeAsO$_{1-x}$F$_x$, both in $c$-axis and in $ab$-planar fields,
by adopting high-quality single crystals.

In this paper, we present $H_{c2}(T)$ for the magnetic fields
along the $c$-axis ($H_{\parallel c}$) and in the $ab$-plane
($H_{\parallel ab}$) for single crystals of oxygen-deficient
SmFeAsO$_{0.85}$ and fluorine-doped SmFeAsO$_{0.8}$F$_{0.2}$.
$H_{c2}(T)$ were determined from the resistive transition in
pulsed (static) magnetic fields up to 60 T (6.9 T). The sublinear
increase of $H_{c2}^{ab}(T)$ with decreasing temperature below
$T_c$, as previously seen in a NdFeAsO$_{0.7}$F$_{0.3}$ single
crystal~\cite{Jaroszynski} and in AEFe$_2$As$_2$
compounds~\cite{Altarawneh,Yamamoto,Baily,Yuan,Kano}, was also
observed in our crystals. On the other hand, $H_{c2}^{c}(T)$'s of
our SmFeAsO$_{0.85}$ and SmFeAsO$_{0.8}$F$_{0.2}$ crystals
linearly increase with decreasing temperature near $T_c$ but tend
to be saturated far below $T_c$. This temperature dependence of
$H_{c2}^{c}$ is in contrast to the linear temperature dependence
found in AEFe$_2$As$_2$ in all the temperature range below
$T_c$~\cite{Altarawneh,Yamamoto,Baily,Yuan,Kano} and to the
significant upturn behavior in NdFeAsO$_{0.7}$F$_{0.3}$ single
crystals below $T_c$~\cite{Jaroszynski}. A deviation of
$H_{c2}(T)$ from the conventional one-gap
Werthamer-Helfand-Hohenberg (WHH) prediction is found in our
crystals. This feature, along with the reduction of its anisotropy
with lowering temperatures, turns out to be common to
iron-pnictides superconductors. Its detailed temperature
dependence and the anisotropy, however, can be quite different
depending on the compounds, which indicates the complex interplay
of a multi-band nature and the paramagnetic effect.

\section{Experiment}
Single crystals of SmFeAsO$_{0.85}$ and SmFeAsO$_{0.8}$F$_{0.2}$
with nominal compositions were grown using the self-flux method
under high-temperature and high-pressure condition. Stoichiometric
starting compounds of SmAs, Fe$_2$O$_3$, and Fe for
SmFeAsO$_{0.85}$ single crystals and SmAs, FeAs, Fe$_2$O$_3$, Fe,
and SmF$_3$ for SmFeAsO$_{0.8}$F$_{0.2}$ single crystals were
used. A pellet sealed in a boron-nitride container was placed in a
cubic pyrophyllite cell equipped with a carbon heater. A 14-mm
cubic multi-anvil-type press was used to pressurize the whole
assembly. Heat treatment at 1350-1450 $^\circ$C was done for 8-10
h under a constant pressure of 3.3 GPa, which was then followed by
rapid cooling to room temperature. Details of the single-crystal
growth are described elsewhere~\cite{HSLee}. After the pressure
was released, the final bulk was mechanically crushed to separate
the single crystals from the flux.

Thus-grown crystals have plate-like shapes. X-ray diffraction
reveals that the crystal surface is normal to the $c$-axis with
the plate-shaped surface along the $ab$-plane. In-plane resistive
transition of SmFeAsO$_{0.85}$ and SmFeAsO$_{0.8}$F$_{0.2}$ single
crystals was measured using the standard four-probe technique.
Contact leads were prepared by using photolithography on the
plate-like sample surface. The upper insets of Figs. 1(a) and (b)
show optical microscopic images of the four-probe patterned
SmFeAsO$_{0.85}$ [dimensions: $\sim$80$\times$50$\times$10
$\mu$m$^3$] and SmFeAsO$_{0.8}$F$_{0.2}$ [dimensions:
$\sim$60$\times$50$\times$10 $\mu$m$^3$] single-crystal specimens,
respectively. The resistive transition [$R(T)$] was measured in
low applied magnetic fields up to 6.9 T. Resistance as a function
of fields [$R(H)$] up to 60 T was also measured at different
temperatures in pulsed-field facilities at Hochfeld-Magnetlabor
Dresden. During the measurements, magnetic fields were applied
along the $c$ axis and in the $ab$ plane while maintaining the
current flow of 1 mA normal to the magnetic field.

\section{Results}
As shown in the lower insets of Figs. 1(a) and (b), the
superconducting transitions in zero field are very sharp for both
SmFeAsO$_{0.85}$ and SmFeAsO$_{0.8}$F$_{0.2}$ crystals. The onset
of the superconducting transition, defined by the deviation from
the linear $R(T)$ above $T_c$, occurs at about $T_{c,onset}$=50.5
K for SmFeAsO$_{0.85}$ and about 42 K for
SmFeAsO$_{0.8}$F$_{0.2}$. The transition width $\Delta T_c$,
determined by adopting the criterion of 10-90\% of the
normal-state resistance $R_n$, is $\sim$ 0.5 K for
SmFeAsO$_{0.85}$ and $\sim$0.8 K for SmFeAsO$_{0.8}$F$_{0.2}$.
$\Delta T_c$'s for both crystals are much narrower than the
reported values of 2$-$4 K for single crystalline
REFeAsO$_{1-x}$F$_x$ (RE = Sm, Nd)~\cite{Karpinski,Jia1,Jia2},
indicating good quality of our samples. As Figs. 1 (a) and (b)
show, the residual resistivity ratio $RRR$$\equiv$$\rho$(300
K)/$\rho(T_{c,onset})$ of $\sim$4.5 for SmFeAsO$_{0.85}$ is larger
than the value of 2.5 seen previously for
NdFeAsO$_{0.82}$F$_{0.18}$ single crystals~\cite{Jia1,Jia2}, while
$RRR$ of SmFeAsO$_{0.8}$F$_{0.2}$ is $\sim$2.2, which is somewhat
smaller but still comparable to that of the previous report. This
indicates that the impurity scattering effect in our fluorine-free
SmFeAsO$_{0.85}$ single crystal is less than the fluorine-doped
REFeAsO$_{1-x}$F$_{x}$ single crystal. According to the recent
report~\cite{Yang}, fluorine does not fully substitute for oxygen
and, thus, some oxygen vacancies remain in the crystal. The
resulting additional scattering centers in REFeAsO$_{1-x}$F$_{x}$
may have enhanced the impurity scattering and led to a smaller
$RRR$ value than in SmFeAsO$_{1-x}$.

Figures 2(a) and (b) present temperature dependence of resistance
[$R(T)$] of SmFeAsO$_{0.85}$ single crystal in low magnetic fields
from 0 to 6.9 T for $H_{\parallel c}$ and $H_{\parallel ab}$,
respectively. The corresponding $R(T)$ of SmFeAsO$_{0.8}$F$_{0.2}$
single crystal is displayed in Figs. 3(a) and (b). Upon increasing
magnetic fields, the resistive transition in $H_{\parallel c}$
becomes broader and the onset of superconductivity shifts to lower
temperatures. The trend is more conspicuous in $H_{\parallel c}$
than in $H_{\parallel ab}$. These behaviors of $R(T)$ in low
magnetic fields of $H_{\parallel c}$ and $H_{\parallel ab}$ in
both crystals are similar to what was previously reported for
SmFeAsO$_{0.7}$F$_{0.25}$~\cite{Karpinski} and
NdFeAsO$_{0.82}$F$_{0.18}$~\cite{Jia1,Jaroszynski} single
crystals. However, details of the magnetic field dependence are
notably different from the previous observation. In our crystals,
the resistive tail is more clearly observed, for $H_{\parallel c}$
in particular, with a gradual extension to lower temperatures with
increasing fields [see Fig. 2(a) for SmFeAsO$_{0.85}$ and Fig.
3(a) for SmFeAsO$_{0.8}$F$_{0.2}$]. The tail of $R(T)$ was also
observed in high $H_{\parallel c}$ in
cuprates~\cite{Pradhan,Safar,Fendrich,Lopez} and in YNi$_2$B$_2$C
~\cite{Mun}, where both have layered structure with CuO$_2$ and
Ni$_2$B$_2$ conducting planes, respectively. It has been known
that such a resistive tail can be explained in terms of the
vortex-glass phase~\cite{Fisher}. Therefore, the observation of
the $R(T)$ tail in our crystals indicates the possible formation
of the vortex-glass phase for $H$$>$2 T. In the same
magnetic-field region of $H_{\parallel c}$, the formation of
vortex-liquid phase was also confirmed in NdFeAsO$_{1-x}$F$_x$
single crystals~\cite{Pribulova}. Oxygen vacancies in both of our
SmFeAsO$_{0.85}$ and SmFeAsO$_{0.8}$F$_{0.2}$ crystals may have
acted as random intrinsic point defects and induced the
vortex-glass state. Detailed analysis on the vortex dynamics in
our crystals will be presented separately~\cite{HSLee1}.

The magnetic-field dependence of resistance [$R(H)$] of our
SmFeAsO$_{0.85}$ [Figs. 2(c)-(d)] and SmFeAsO$_{0.8}$F$_{0.2}$
[Figs. 3(c)-(d)] crystals was measured in pulsed magnetic fields
(both $H_{\parallel c}$ and $H_{\parallel ab}$) up to 60 T at
various temperatures. The upper critical fields, $H_{c2}^c$ for
$H_{\parallel c}$ and $H_{c2}^{ab}$ for $H_{\parallel ab}$, were
obtained by adopting different criteria; 90\%, 50\%, and 10\% of
$R_n$. The normal-state resistance $R_n$ was determined by
linearly extrapolating the normal-state behavior above the onset
of superconductivity in $R(T)$ and $R(H)$ curves separately.
Thus-determined values of $H_{c2}^c$ and $H_{c2}^{ab}$ are shown
in Figs. 4(a) and (b) for SmFeAsO$_{0.85}$ [in Figs. 4(c) and (d)
for SmFeAsO$_{0.8}$F$_{0.2}$]. In both crystals, $H_{c2}(T)$
obtained from $R(T)$ (in Figs. 2 and 3) at low static magnetic
fields (open symbols) is in line with those from $R(H)$ curves (in
Figs. 2 and 3) at high pulsed magnetic fields (solid symbols).
With lowering temperature, $H_{c2}^c(T)$ exhibits a slight upturn
variation for the 10\% criteria but it turns gradually into a
slight downturn curvature as one moves to 90\% criteria, in
particular for SmFeAsO$_{0.8}$F$_{0.2}$ sample. Similar behavior
has also been observed in
cuprates~\cite{Ando,Welp,Fournier,Sumarlin,Vedeneev1}. The
criteria-dependent discrepancy arises from the fact that the
region near 10\% of $R_n$ is related to the vortex-liquid phase
while the region near 90\% of $R_n$ is affected by the
superconducting
fluctuation~\cite{Welp,Fournier,Sumarlin,Vedeneev1}. Thus, we
adopt the 50\%-$R_n$ criterion to determine $H_{c2}(T)$. The
resultant $H_{c2}(T)$ for $H_{\parallel c}$ and $H_{\parallel ab}$
is summarized in Fig. 5(a) for SmFeAsO$_{0.85}$ [in Fig. 5(b) for
SmFeAsO$_{0.8}$F$_{0.2}$]. The values of $H_{c2}(T)$ for
$H_{\parallel c}$ and $H_{\parallel ab}$ of our SmFeAsO$_{0.85}$
and SmFeAsO$_{0.8}$F$_{0.2}$ single crystals are in the same range
as reported for the corresponding polycrystals, with the similar
value of $T_c$$\sim$50 K and $\sim$40 K, respectively, for the two
crystals~\cite{Jo,Jaroszynski1}.

According to a recent report on high-field resistivity
measurements in a NdFeAsO$_{0.7}$F$_{0.3}$ single crystal up to 60
T~\cite{Jaroszynski}, $H_{c2}(T)$ for $H_{\parallel c}$ exhibits a
pronounced upturn in the entire ranges of magnetic field 60 T and
temperature below $T_c$$\sim$45 K. This result is similar to the
earlier report for polycrystalline samples~\cite{Hunte,Jo}, where
the upturn shape of $H_{c2}^c(T)$ was suggested to be an intrinsic
property of iron pnictides and was explained in terms of the
two-band model~\cite{Jaroszynski,Hunte,Jo}. However, our
SmFeAsO$_{0.85}$ and SmFeAsO$_{0.8}$F$_{0.2}$ single crystals show
linear increase of $H_{c2}^c(T)$ with decreasing temperature near
$T_c$ but tend to be saturated far below $T_c$ [see Figs. 5(a) and
(b)]. This temperature dependence of $H_{c2}^{c}$ is in contrast
to the linear and upturn temperature dependences in AEFe$_2$As$_2$
~\cite{Altarawneh,Yamamoto,Baily,Yuan,Kano} and in
NdFeAsO$_{0.7}$F$_{0.3}$ single crystals~\cite{Jaroszynski}, respectively, below $T_c$.

\section{Discussion}
First, we compare the $H_{c2}(T)$ data of our crystals with the
conventional WHH theory~\cite{WHH}, which is based on the orbital
effect arising from the Lorentz force acting on paired electrons
with opposite momenta as the main cause of pair breaking. In
addition, the theory is extended to include the effects of spin
paramagnetism ($\alpha$) and spin-orbit scattering
($\lambda_{so}$). Here, we assume that the spin-orbit scattering
due to impurities is negligible ($\lambda_{so}$=0)~\cite{Fuchs}.
As shown in the Fig. 5, the data points of $H_{c2}(T)$ for
$H_{\parallel c}$ and $H_{\parallel ab}$ in both crystals do not
well follow the WHH model for $\alpha$=0 (solid lines).
$H_{c2}^c(T)$'s for both crystals are enhanced compared to the WHH
prediction, while $H_{c2}^{ab}(T)$'s are suppressed below the WHH
curve with a flattening behavior.

Using the WHH theory for $\alpha$=0 and $\lambda_{so}$=0, we
estimate the $H_{c2}$ value at $T$=0 [$H_{c2,WHH}$(0)].
$H_{c2,WHH}^{c}$(0)$\approx$84 T [47 T] and
$H_{c2,WHH}^{ab}$(0)$\approx$378 T [280 T] for SmFeAsO$_{0.85}$
[SmFeAsO$_{0.8}$F$_{0.2}$] are obtained using the relation,
$H_{c2,WHH}$(0)=0.69$T_c |dH_{c2}/dT|_{T_c}$. The values of
$|dH_{c2}/dT|_{T_c}$ are presented in Table~\ref{tab:table1}. It
is noteworthy to compare these $H_{c2,WHH}$(0) values with the
paramagnetic limiting field due to Zeeman paramagnetic pair
breaking, $H_p$($T$=0)$\cong$(1+$\lambda_{ep}$)$H_p^{BCS}$($T$=0).
Here, $H_p^{BCS}$($T$=0)=1.84$T_c$($H$=0)~\cite{Clogston} is the
Pauli or Clongston-Chandrasekhar-limit field for isotropic
$s$-wave pairing in the absence of spin-orbit scattering in weakly
coupled superconductors. $\lambda_{ep}$ is introduced to take into
account the strong electron-boson ($i.e.$, phonon) coupling in the
system. If we take~\cite{Jaroszynski,Fuchs,Drechsler}
$\lambda_{ep}=0.6$, $H_p$(0)'s for SmFeAsO$_{0.85}$ and
SmFeAsO$_{0.8}$F$_{0.2}$ are estimated to be about 145 T and 120
T, respectively. In both crystals, the values of
$H_{c2,WHH}^{c}$(0) are much smaller than the corresponding values
of $H_p$(0). This indicates that the $H_{c2}^c$ is determined
dominantly by the orbital effect rather than the paramagnetic
effect. By contrast, $H_{c2,WHH}^{ab}$(0) is much larger than
$H_p$(0). In fact, the $H_{c2}^{ab}(T)$'s of both crystals have a
tendency to be suppressed below the WHH curve for $\alpha$=0 and
thus the actual $H_{c2}^{ab}(0)$ is expected to be much smaller
than $H_{c2,WHH}^{ab}$(0) estimated based on the paramagnetic
effect. For $H_{c2}^c(T)$ where the orbital effect is dominant, on
the other hand, its enhancement compared to the WHH curve with
$\alpha$=0 cannot be explained in terms of the conventional
one-gap WHH theory.

In order to understand the detailed temperature dependence of
$H_{c2}^c(T)$, we consider the multi-band nature of
iron-pnictides. It has been well-known that there are two
different coexisting groups of Fermi surfaces: one with electron
and the other with hole
character~\cite{Mazin,Yao,Li1,Marsiglio,Han}. Using the two-gap
dirty-limit model of $H_{c2}(T)$~\cite{Gurevich} we can fit the
experimental data as shown in Figs. 6(a) and (b). The equation of
$H_{c2}(T)$ for $H_{\parallel c}$ considering orbital pair
breaking is given by $a_0[$ln$ t+U(h)][$ln$ t+U(\eta h)]+a_2[$ln$
t+U(\eta h)] +a_1[$ln$ t+U(h)]=0$, where $t=T/T_c$,
$U(x)=\Psi(1/2+x)-\Psi(x)$, $\Psi(x)$ is the Euler digamma
function, $\eta=D_2/D_1$, $D_{1,2}$ are diffusivities of the bands
1 and 2, and $h=H_{c2}D_1/(2\phi_0T)$. $a_{0,1,2}$ are constants
described with intraband- and interband- coupling constants
$\lambda_{11,22}$ and $\lambda_{12,21}$ in the bands 1 and 2,
respectively. Precise definitions of $a_{0,1,2}$ can be found in
Ref.~\cite{Gurevich}. The equation of $H_{c2}(T)$ can be
generalized to the case of a field inclined by angle $\theta$ with
respect to the $ab$ plane by adopting angle-dependent
diffusivities, $D_{1,2}(\theta)=[(D_{1,2}^{ab})^2$cos$^2\theta+
D_{1,2}^{ab}D_{1,2}^{c}$sin$^2\theta]^{1/2}$~\cite{Gurevich}.
Therefore, $D_{1,2}$ are given by $D_{1,2}^{ab}$ for $H_{\parallel
c}$ and $[D_{1,2}^{ab}D_{1,2}^{c}]^{1/2}$ for $H_{\parallel ab}$,
where $D_{1,2}^{ab}$ ($D_{1,2}^{c}$) are the in-plane
(out-of-plane) electron diffusivities of the bands 1 and 2. We
discuss two different cases; (1) dominant intraband coupling
\emph{w}$>$0 and (2) dominant interband coupling \emph{w}$<$0,
where $w=\lambda_{11}\lambda_{22}-\lambda_{12}\lambda_{21}$. Here,
we take three sets of $\lambda$ for \emph{w}$>$0
[(1)$\lambda_{11}$=0.8, $\lambda_{22}$=0.3,
$\lambda_{12,21}$=0.18, (2)$\lambda_{11,22}$=0.5,
$\lambda_{12,21}$=0.25, (3)$\lambda_{11,22}$=0.7,
$\lambda_{12,21}$=0.5
] and two sets of $\lambda$ for \emph{w}$<$0
[(4)$\lambda_{11,22}$=0.49, $\lambda_{12,21}$=0.5,
(5)$\lambda_{11,22}$=0.5, $\lambda_{12,21}$=0.55]. Due to the lack
of microscopic theory of pairing mechanism, we choose the values
of $\lambda$ close to the ones adopted in earlier
reports~\cite{Jaroszynski,Hunte,Baily}.

First, we consider the case of $H_{\parallel c}$. As shown in Fig.
6 (a) for SmFeAsO$_{0.85}$, the $H_{c2}(T)$ predicted by the
two-gap theory can reproduce nicely the experimental data taken up
to 60 T for all cases. Depending on the sign of \emph{w}, however,
the theoretical curves have different curvatures beyond the field
range of measurements. Near $T$=0, the $H_{c2}$ curves saturate to
the values of $H_{c2}^c(0)$$\sim$110 T for (1) and $\sim$135$-$142
T for (2) and (3) in \emph{w}$>$0, but still rapidly increase with
upturn curvatures toward $H_{c2}^{c}(0)$$\sim$220$-$300 T for (4)
and (5) in \emph{w}$<$0. In these cases, $\eta_{\parallel
c}$=$D_{2}^{ab}/D_{1}^{ab}$ is in the range of $\sim$5-9 for
\emph{w}$>$0 and $\sim$19$-$36 for \emph{w}$<$0. In contrast, for
SmFeAsO$_{0.8}$F$_{0.2}$, the different sets of fitting parameters
lead to almost identical curves, well fitting the $H_{c2}(T)$ data
[Fig. 6(b)] with $H_{c2}^c(0)$$\sim$50 T,
$H_{c2}^{ab}(0)$$\sim$208 T, and $\eta_{\parallel
c}$$\sim$2.2$-$3.7. This indeterminacy of the sign of \emph{w} for
a better fit to the $H_{c2}^c$ of SmFeAsO$_{0.8}$F$_{0.2}$ may
stem from the higher inhomogeneity of the crystal.

If we take into account the difference in the average Fermi
velocities~\cite{Singh1} of hole and electron sheets, the
difference in the intraband diffusivities $D_{2}^{ab}\sim
(19$-$36)D_{1}^{ab}$ for \emph{w}$<$0 looks too high. In addition,
since $H_p$(0)$\sim$145 T was estimated for SmFeAsO$_{0.85}$, the
parameter set of (1) for \emph{w}$>$0, which gives
$H_{c2}^c(0)$$\sim$ 110 T, is more reasonable to explain the
experimental data. Thus, in the reasonable range of
$\eta_{\parallel c}$, the $H_{c2}^{c}(T)$ curves of both crystals
do not show the pronounced upturn behavior in the whole field and
temperature range. This is somewhat different from the
result~\cite{Jaroszynski} of NdFeAsO$_{0.7}$F$_{0.3}$ single
crystal, showing the significant upturn of $H_{c2}^{c}(T)$ at the
field up to 60 T. In Ref.~\cite{Jaroszynski}, such a pronounced
upturn of $H_{c2}^{c}(T)$ is explained in terms of a two-band
model, assuming a large difference in $D_{2}^{ab}$ and
$D_{1}^{ab}$ with $\eta_{\parallel c}$ $\sim$ 10-100. It is not
clear yet whether such a huge difference in $D_{2}^{ab}$ and
$D_{1}^{ab}$ is intrinsic. As pointed out in
Ref.~\cite{Jaroszynski} the strong upturn in $H_{c2}^{c}(T)$ can
be due to scatterings at magnetic impurities. In any case, the
strong deviation from the WHH model is a common feature of
iron-pnictides, which reflects the multi-gap nature of the
materials.

Next, we consider the case of $H_{\parallel ab}$. In both
crystals, the various sets of fitting parameters using the two-gap
model lead to an identical curve of $H_{c2}^{ab}(T)$. As shown in
Figs. 6(a) and (b), the fitting curves of two-band model cannot
capture the flattening behavior with decreasing temperatures in
both crystals. As discussed above, for $H_{\parallel ab}$ we
expect that the paramagnetic limiting plays an essential role for
determining $H_{c2}^{ab}(T)$. In the framework of the WHH theory,
such a spin-paramagnetic effect can be taken into account by
introducing the so-called Maki parameter, $\alpha$. With
$\alpha$=2.3 and 2.7, the $H_{c2}^{ab}(T)$ data of
SmFeAsO$_{0.85}$ and SmFeAsO$_{0.8}$F$_{0.2}$ crystals are nicely
fitted by the WHH model [see Fig. 7]. It has been known that the
Maki parameter $\alpha$ becomes larger as the system is
disordered~\cite{Fuchs}. A slightly larger value of $\alpha$ for
SmFeAsO$_{0.8}$F$_{0.2}$ than for SmFeAsO$_{0.85}$ is consistent
with its smaller $RRR$ value. The values of $H_{c2,WHH}^{p}$(0)
obtained by considering the Pauli paramagnetism with $\alpha\neq$0
in the WHH theory are estimated to be $\approx$150 T for
SmFeAsO$_{0.85}$ and $\approx$100 T for SmFeAsO$_{0.8}$F$_{0.2}$.
Since $H_{c2,WHH}^{p}(0)\geq H_p(0)$ in our crystals, the data of
$H_{c2}^{ab}(T)$ are strongly affected by the spin paramagnetic
effect rather than by the two-band nature.

\begin{table*}
\caption{\label{tab:table1} Superconducting parameters of
SmFeAsO$_{0.85}$ and SmFeAsO$_{0.8}$F$_{0.2}$ single crystals
obtained from the analysis of $H_{c2}(T)$. The $c$-axis and the
$ab$-plane coherence length, $\xi_{c}(0)$ and $\xi_{ab}(0)$,
respectively, are estimated with the Ginzburg-Landau relations for
the upper critical field of
$H_{c2}^c$=$\Phi_0$/$2\pi\xi_{ab}^2(0)$ and
$H_{c2}^{ab}$=$\Phi_0$/$2\pi\xi_{ab}(0)\xi_{c}(0)$.}
\begin{ruledtabular}
\begin{tabular}{cccccccc}
&$T_{c,onset}$ & $|\frac{dH_{c2}}{dT}|_{T_{c}//c}$ &
$|\frac{dH_{c2}}{dT}|_{T_{c}//ab}$& $H_{c2}^c$(0)\footnotemark[4]
& $H_{c2}^{ab}$(0)\footnotemark[5] & $\xi_{ab}(0)$ & $\xi_{c}(0)$ \\
& [K] & [T/K] & [T/K] & [T] & [T] & [{\AA}] & [{\AA}]\\
\hline
SmFeAsO$_{0.85}$ & 50.5 & 2.5 & 11 & 110 & 150 & 17 & 3.6 \\
SmFeAsO$_{0.8}$F$_{0.2}$ & 42 & 1.7 & 9.9 & 50 & 100 & 26 & 3.6 \\
\end{tabular}
\end{ruledtabular}
\footnotetext[4]{$H_{c2}^c(0)$ is determined from the analysis
with two-band model.} \footnotetext[5]{$H_{c2}^{ab}(0)\equiv
H_{c2,WHH}^p(T=0)$ is estimated with the WHH theory including
paramagnetism. }
\end{table*}
Due to the quasi-two-dimensional Fermi-surface topology, for a
$H_{\parallel c}$, the cross-section of the Fermi-surface produces
closed current loops that form
vortices~\cite{Shrieffer,Vedeneev2,Li2,Singleton1,Nam}. Thus, for
$H_{\parallel c}$, the orbital pair-breaking mechanism plays a
dominant role in destroying the superconductivity in high magnetic
fields. Thus, the two-gap theory, taking into account the orbital
pair-breaking effect, well describes our $H_{c2}^c(T)$ data. For a
$H_{\parallel ab}$, however, closed loops cannot be easily formed
because the cross-sectional area of the Fermi-surface is almost
fully open~\cite{Singleton2} with negligible orbital effect, thus
resulting in a rapid increase of $H_{c2}(T)$ near $T_c$.
Therefore, the spin-paramagnetic effect is a more dominant factor
in reducing the increase rate of $H_{c2}^{ab}(T)$ with decreasing
temperature in our crystals.

In Figure 8, the anisotropy of $H_{c2}$,
$\gamma$$\equiv$$H_{c2}^{ab}$/$H_{c2}^c$, is plotted as a function
of reduced temperature $t$=$T/T_c$ for SmFeAsO$_{0.85}$ (circles)
and SmFeAsO$_{0.8}$F$_{0.2}$ (diamonds). The value of $\gamma$ for
SmFeAsO$_{0.85}$ (SmFeAsO$_{0.8}$F$_{0.2}$) crystal is in the
range of about 3$-$6 (4$-$7), at the temperature region of
$T$=(0.75$-$1)$T_c$. SmFeAsO$_{0.8}$F$_{0.2}$ has a somewhat
larger $\gamma$ than SmFeAsO$_{0.85}$. These values are similar in
magnitude to the ones reported in other REFeAsO$_{1-x}$F$_x$ (RE =
Sm and Nd) single
crystals~\cite{Karpinski,Jia1,Jia2,Jaroszynski,Weyeneth1,Weyeneth2}.
The $\gamma$ has temperature dependence, which is distinct from
that of the conventional single-band superconductivity. The
decreasing $\gamma$ with decreasing temperature in both crystals
results from the enhanced $H_{c2}^c(T)$ and the suppressed
$H_{c2}^{ab}(T)$ compared to the WHH for $\alpha$=0 as shown in
Figs. 5 (a) and (b). Therefore, the temperature dependence of
$\gamma$ originates from the combined effect of two-band nature
and spin paramagnetism.

\section{Summary}
This study reports on $H_{c2}^c(T)$ and $H_{c2}^{ab}(T)$ of
fluorine-free SmFeAsO$_{0.85}$ and fluorine-doped
SmFeAsO$_{0.8}$F$_{0.2}$ single crystals, investigated by
measuring the resistive transition at high magnetic fields up to
60 T. In contrast to the strong upturn curvature of
$H_{c2}^{c}(T)$ reported earlier in NdFeAsO$_{1-x}$F$_x$ single
crystal, $H_{c2}^{c}(T)$'s in both of our crystals increase
linearly with decreasing temperature near $T_c$ and tends to be
saturated at low-enough temperatures. We confirm that the
temperature dependences of $H_{c2}^{c}(T)$'s well follow the
two-gap dirty-limit prediction while they deviate from the one-gap
WHH prediction, regardless of inclusion of the spin-paramagnetic
effect. On the other hand, $H_{c2}^{ab}(T)$ of our crystals show
the downturn curvature, consistent with the earlier observation in
NdFeAsO$_{1-x}$F$_x$ single crystal and AEFe$_2$As$_2$ compounds.
The importance of paramagnetic effect on the downturn curvature in
$H_{c2}^{ab}(T)$ has already been pointed out for
NdFeAsO$_{1-x}$F$_x$ single crystal and AEFe$_2$As$_2$ compounds,
but $H_{c2}^{ab}(T)$ data were analyzed only within the two-gap
model~\cite{Jaroszynski,Baily,Yuan,Kano}. In this study, the
temperature dependences of $H_{c2}^{ab}(T)$'s are analyzed in
terms of two-gap model and the WHH theory including the
paramagnetic effect. Our analysis clearly indicates that the
flattening of $H_{c2}^{ab}(T)$ is governed mainly by the
paramagnetic pair-breaking effect rather than the two-gap effect.
This study shows that the upper critical field in Sm-based
iron-pnictides is determined by the complex interplay of a
two-band nature and the paramagnetic effect depending on the
direction of magnetic field application with respect to the
crystal axes. We believe this is the generic characteristics of
different families of iron-pnictide compounds.

\begin{acknowledgments}
This work was supported by the Korea Science and Engineering
Foundation through Acceleration Research Grant
R17-2008-007-01001-0 and by POSCO. JSK is supported by the POSTECH
Research funding.
\end{acknowledgments}

\newpage
\begin{figure}
\includegraphics[width=8cm]{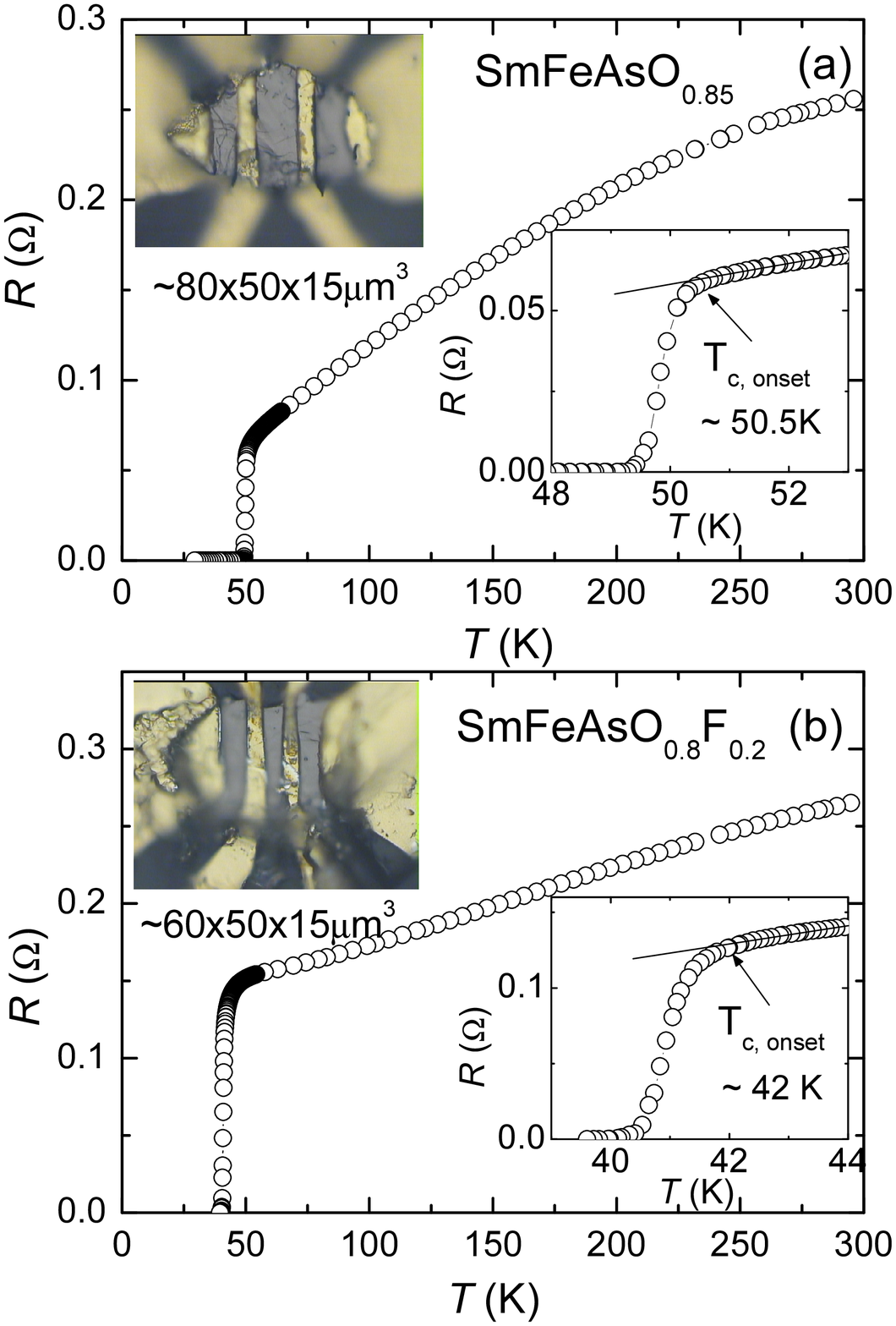}
\caption{Temperature dependence of the resistance $R(T)$ of (a)
SmFeAsO$_{0.85}$ and (b) SmFeAsO$_{0.8}$F$_{0.2}$ single crystals
in zero magnetic field. The upper insets of (a) and (b) show the
microscopic images of the four-probe patterned crystals used for
the transport measurements. The lower insets of (a) and (b) show a
magnified view near the superconducting transition. The onset of
superconducting transition, $T_{c,onset}$, defined by the
deviation from the linearity of $R(T)$, is indicated in the lower
insets. } \label{fig1}
\end{figure}
\newpage
\begin{figure}
\includegraphics[width=8cm]{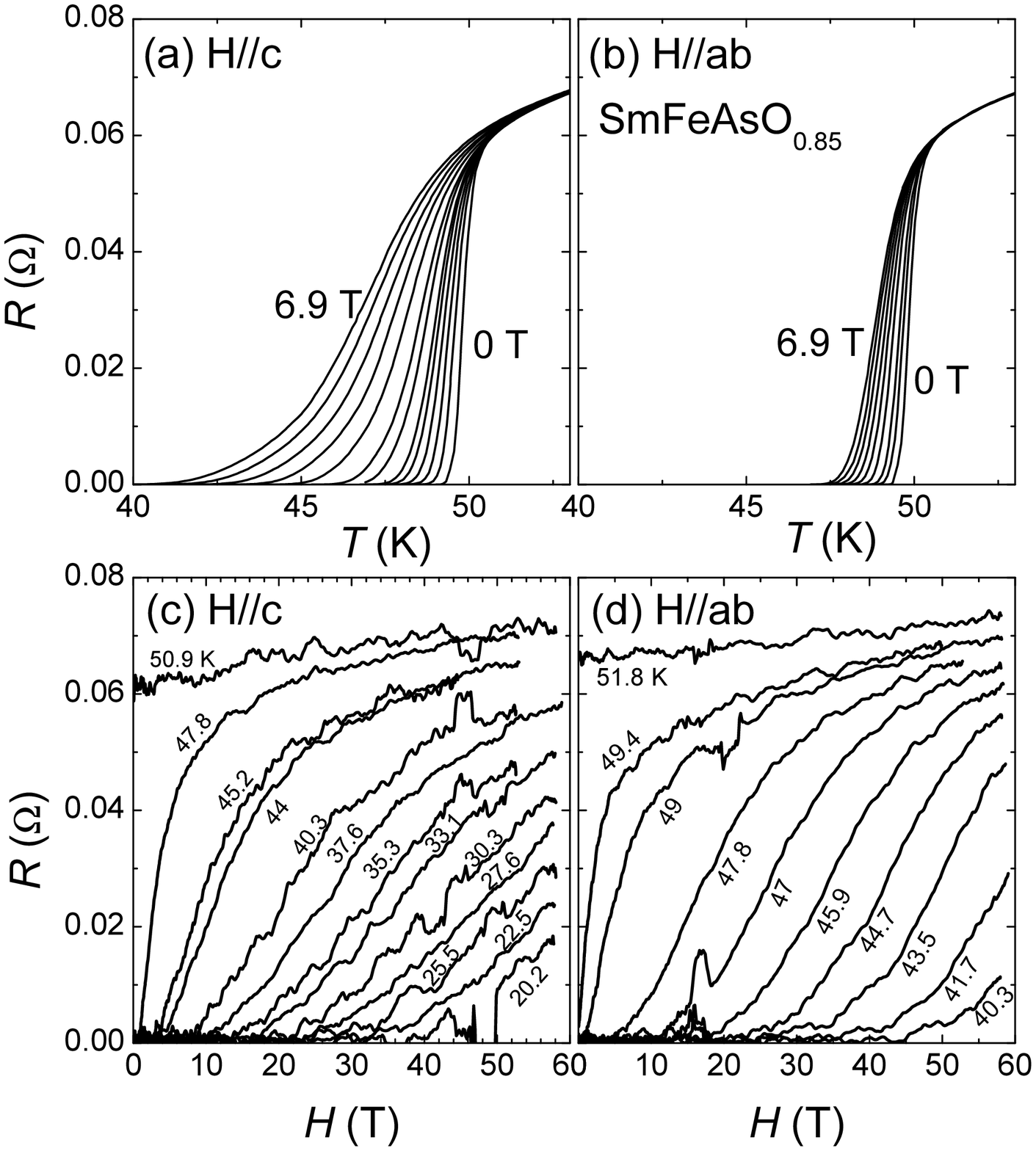}
\caption{Temperature dependence of the resistance $R(T)$ of
SmFeAsO$_{0.85}$ single crystal measured at the various static
fields from 0 to 6.9 T for (a) $H_{\parallel c}$ (0, 0.2, 0.4,
0.6, 0.8, 1, 1.5, 2, 3, 4, 5, 6, 6.9 T) and (b) $H_{\parallel ab}$
(0, 0.5, 1, 2, 3, 4, 5, 6, 6.9 T). Field dependence of the
resistance $R(H)$ of SmFeAsO$_{0.85}$ single crystal measured at
various temperatures in pulsed magnetic fields up to 60 T for (c)
$H_{\parallel c}$ and (d) $H_{\parallel ab}$.} \label{fig2}
\end{figure}
\newpage
\begin{figure}
\includegraphics[width=8cm]{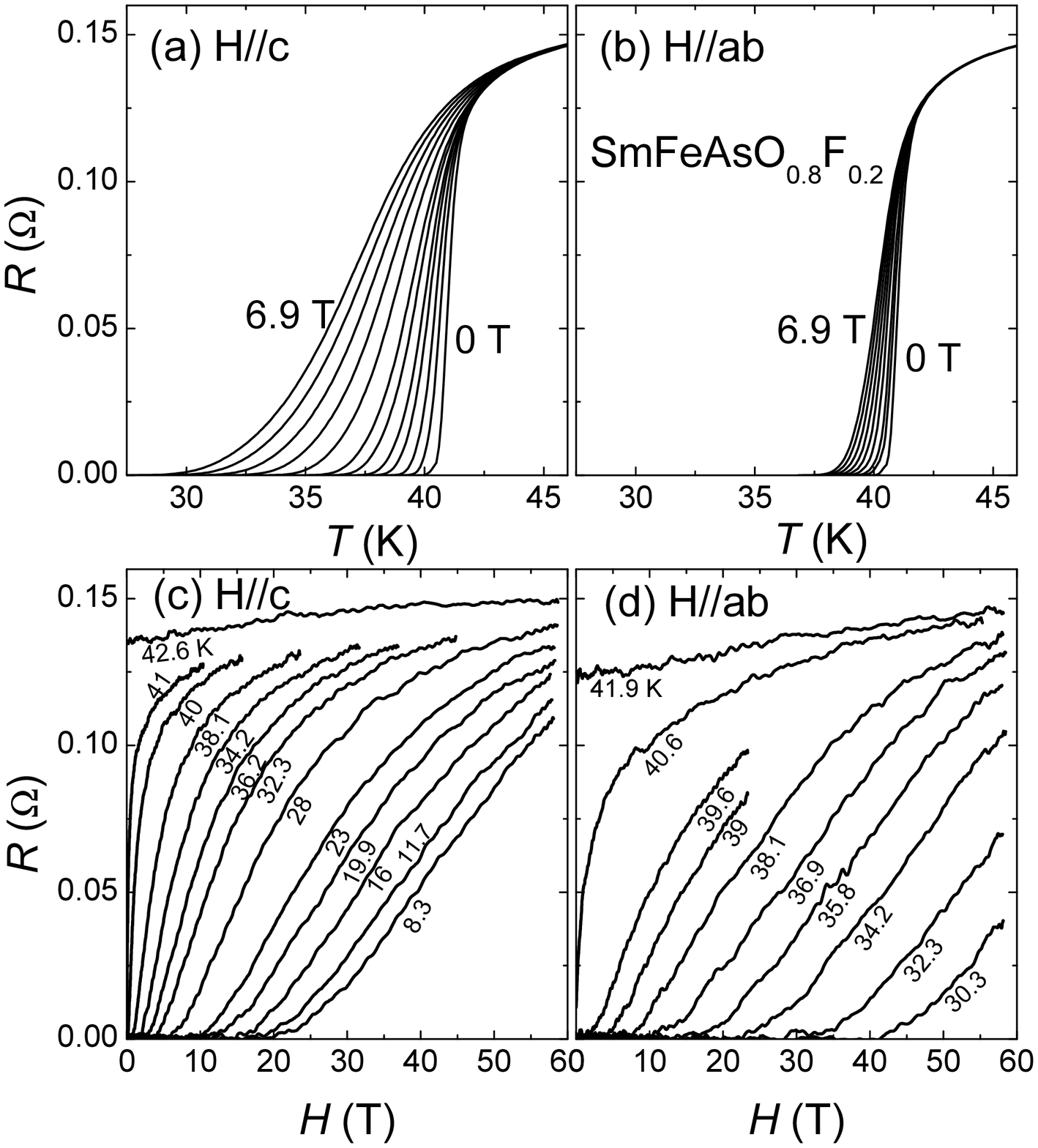}
\caption{Temperature dependence of the resistance $R(T)$ of
SmFeAsO$_{0.8}$F$_{0.2}$ single crystal measured in the various
static fields from 0 to 6.9 T for (a) $H_{\parallel c}$ (0, 0.2,
0.4, 0.7, 1, 1.5, 2, 3, 4, 5, 6, 6.9 T) and (b) $H_{\parallel ab}$
(0, 0.5, 1, 2, 3, 4, 5, 6, 6.9 T). Field dependence of the
resistance $R(H)$ of SmFeAsO$_{0.8}$F$_{0.2}$ single crystal
measured at various temperatures in pulsed magnetic fields up to
60 T for (c) $H_{\parallel c}$ and (d) $H_{\parallel ab}$.}
\label{fig3}
\end{figure}
\newpage
\begin{figure}
\includegraphics[width=8cm]{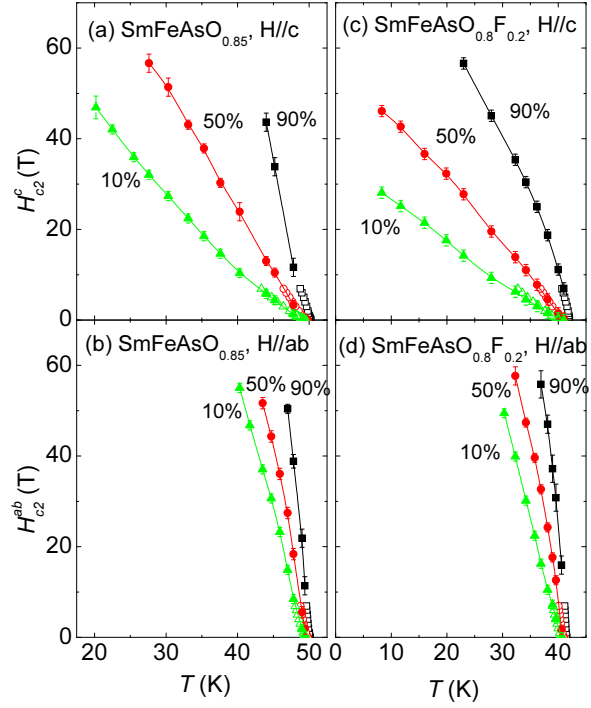}
\caption{(color online) Temperature dependence of the upper
critical fields $H_{c2}(T)$ of SmFeAsO$_{0.85}$ for (a)
$H_{\parallel c}$ and (b) $H_{\parallel ab}$ and of
SmFeAsO$_{0.8}$F$_{0.2}$ for (c) $H_{\parallel c}$ and (d)
$H_{\parallel ab}$. $H_{c2}(T)$ values are extracted from 10\%,
50\%, and 90\% of the normal-state resistance $R_n$ determined by
the linear extrapolation above the onset of superconductivity of
$R(T)$ (open symbols) and of $R(H)$ (closed symbols) curves.}
\label{fig4}
\end{figure}
\newpage
\begin{figure}
\includegraphics[width=7cm]{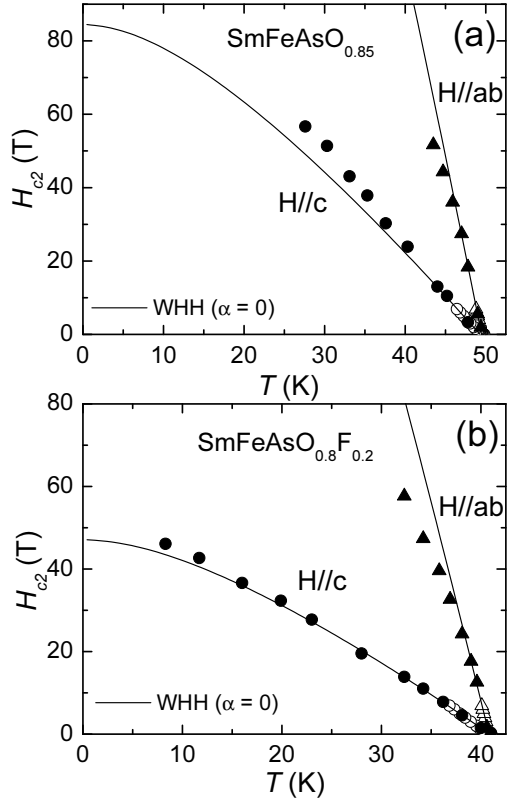}
\caption{$H_{c2}(T)$ for $H_{\parallel c}$ (circles) and
$H_{\parallel ab}$ (triangles) of (a) SmFeAsO$_{0.85}$ and (b)
SmFeAsO$_{0.8}$F$_{0.2}$ extracted from 50\% of $R_n$ shown in
Figs. 4(a)-(d). The open and closed symbols indicate the values of
$H_{c2}(T)$ obtained from $R(T)$ and $R(H)$ curves, respectively.
The experimental data were analyzed in terms of the WHH theory,
which is exhibited by solid lines without the spin-paramagnetic
effect ($\alpha$=0).} \label{fig5}
\end{figure}
\newpage
\begin{figure}
\includegraphics[width=8cm]{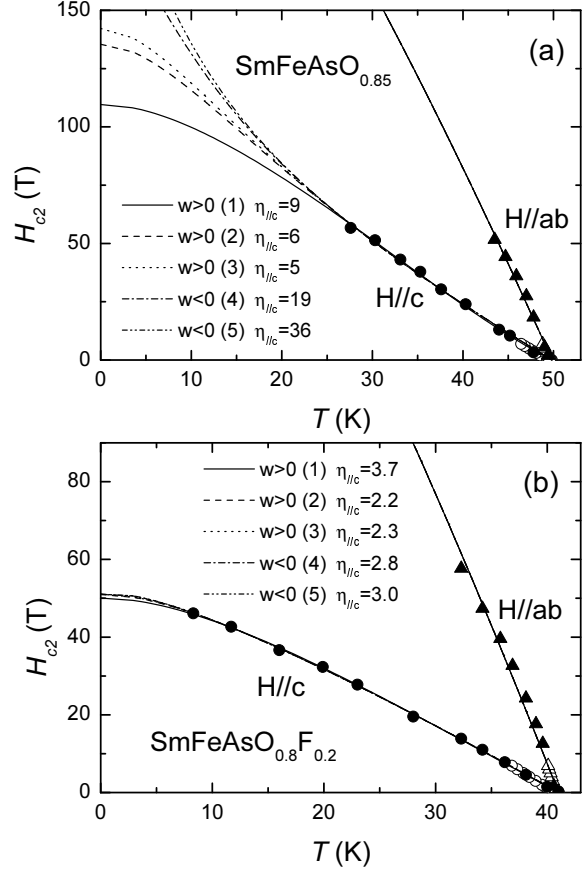}
\caption{Analysis of $H_{c2}(T)$ for $H_{\parallel c}$ and
$H_{\parallel ab}$ of (a) SmFeAsO$_{0.85}$ and (b)
SmFeAsO$_{0.8}$F$_{0.2}$ single crystals using the two-gap theory
in different pairing scenario: for \emph{w}$>$0, (1)
$\lambda_{11}$=0.8, $\lambda_{22}$=0.3, $\lambda_{12,21}$=0.18,
(2) $\lambda_{11,22}$=0.5, $\lambda_{12,21}$=0.25, and (3)
$\lambda_{11,22}$=0.7, $\lambda_{12,21}$=0.5), and for
\emph{w}$<$0, (4) $\lambda_{11,22}$=0.49, $\lambda_{12,21}$=0.5,
and (5) $\lambda_{11,22}$=0.5, $\lambda_{12,21}$=0.55, where
$w=\lambda_{11}\lambda_{22}-\lambda_{12}\lambda_{21}$. The best
fit is obtained for different values of $\eta_{\parallel
c}$=$D_{2}^{ab}/D_{1}^{ab}$ in the inset and an identical value of
$\eta_{\parallel
ab}$=$[D_{2}^{ab}D_{2}^{c}/D_{1}^{ab}D_{1}^{c}]^{1/2}$=1.}
\label{fig6}
\end{figure}
\newpage\begin{figure}
\includegraphics[width=8cm]{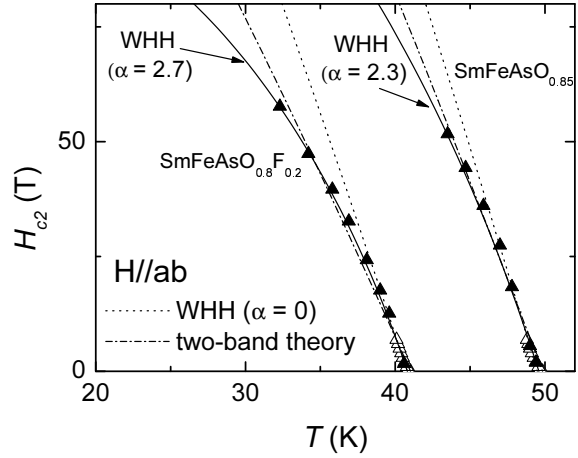}
\caption{Analysis of $H_{c2}(T)$ for $H_{\parallel ab}$ by using
the WHH theory including spin-paramagnetic effect ($\alpha\neq$0).
The flattening of $H_{c2}(T)$ for $H_{\parallel ab}$ of
SmFeAsO$_{0.85}$ and SmFeAsO$_{0.8}$F$_{0.2}$ single crystals,
which can be clearly observed by a deviation from the dotted lines
of the WHH theory for $\alpha$=0, is well explained by the WHH
theory for $\alpha$=2.3 and 2.7 (solid lines), respectively,
rather than by the two-band model (dash-dotted lines).}
\label{fig7}
\end{figure}
\newpage
\begin{figure}
\includegraphics[width=7.5cm]{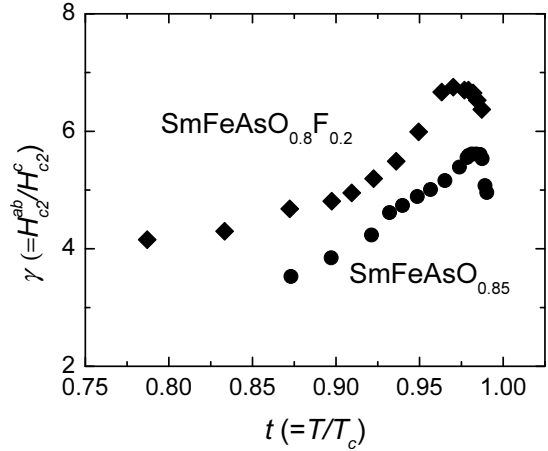}
\caption{Anisotropy ratio of $H_{c2}$,
$\gamma$$\equiv$$H_{c2}^{ab}$/$H_{c2}^c$, as a function of reduced
temperature $t$=$T/T_c$ for SmFeAsO$_{0.85}$ (circles) and
SmFeAsO$_{0.8}$F$_{0.2}$ (diamonds) single crystals.} \label{fig8}
\end{figure}
\end{document}